# Update on the method using nonlocal/noncontact differential conductance experimental data for direct probing Andreev reflections and for extracting the superconducting gap

Nadina Gheorghiu

**Abstract**

In our previous work, high-temperature superconductivity (HTSC) was probed for the first time by using electrical differential conductance $G_{diff}(V) = dI/dV$ data as obtained from current-voltage $V(I)$ measurements on a hydrogenated graphitic fiber [1]. While our nonlocal method for finding the superconducting (SC) gap was recently applied to a conventional SC (Pb [2]), the nonlocal $G_{diff}$ method was first proposed in [1]. Herein, we are bringing forth an update on the topic.

**Earlier, newer results, and their interpretation**

While local, i.e. at the atomic level, $G_{diff}(V)$ measurements are usually done using scanning tunneling spectroscopy, the highly disordered nature of an unconventional material like hydrogenated graphitic material suggests instead the need for using nonlocal $G_{diff}$ measurements. Unlike local electrical conductance spectroscopy, nonlocal $G_{diff}$ measurements can distinguish between nontopological zero-energy modes that are localized around potential inhomogeneities and true Majorana edge modes of the topological phase [3]. When normal metal leads are connected to a SC, two processes are at the core of nonlocal response in $G_{diff}$: 1) direct electron transfer between the normal leads; 2) crossed Andreev reflection of an electron from one lead into another lead. The height of the peak in $G_{diff}(V)$ is proportional to the nonlocal density of states (DOS) of the material. The gap in the energy spectrum is a measure of correlations between electrons, in particular Cooper pair correlations in a SC. At the same time, nonlocal $G_{diff}$ measurements can reveal the presence of Andreev reflections, in which an incoming electron is converted into a reflected hole while a Cooper pair of electrons is transmitted. That is because a weak link between two SC grains with different orientations hosts discrete, fermionic modes otherwise known as Andreev levels [4] that are acting in unison thus giving rise to macroscopic phenomena such as the Josephson SC current. Because each Andreev level is itself a fermionic degree of freedom, it can be populated by electronic excitations known as Bogoliubov quasiparticles. In our previous work, we have revealed the existence of both Andreev reflections and Bogoliubov excitations in oxygen-implanted carbon (C)-based materials [5,6]. While the first coherent manipulation [7] of an Andreev spin qubit was recently demonstrated [8], a graphene qubit [9] is a challenge for material science and bioethics as well.

In [1], we have introduced the analysis of the nonlocal differential conductance $G_{diff}$ obtained by differentiating the $I(V)$ experimental data as a straightforward method of determining the SC gap for hydrogenated polyacrylonitrile (($CH_2$-CH-CN$)_n$ fibers (H-C-N fibers). The results are shown in Fig. 1 (Fig. 4 in [1]). One remarkable feature is the asymmetry in $G_{diff}(V)$. When the particle-hole and the time reversal symmetries are violated, the differential tunneling (local) electrical conductivity and the dynamic (nonlocal) $G_{diff}$ are no more symmetric function of applied voltage $V$ [10]. This asymmetry can be observed both in the normal and SC phases of strongly correlated systems. The conductivity asymmetry is not observed in conventional metals like Pb or Sn, especially at low $T$. In fact, the asymmetry in the tunneling conductance feature is an important sign of the underlying Mott character in doped insulating systems [11] that can show unconventional SC. The asymmetry in $G_{diff}(V)$ mirrors the interference of chiral Andreev edge states, which is a topological phenomenon. In addition, the observed negative $G_{diff}$ (Fig. 2 or Fig. 4b in [1]) is attributed to the nonlocal coherence between electron and holes in the Andreev edge states. Thus, the nonlocal negative conductance/resistance indicates the presence of crossed Andreev converted holes, namely the coupling between two quantum Hall edge states via a narrow SC link.

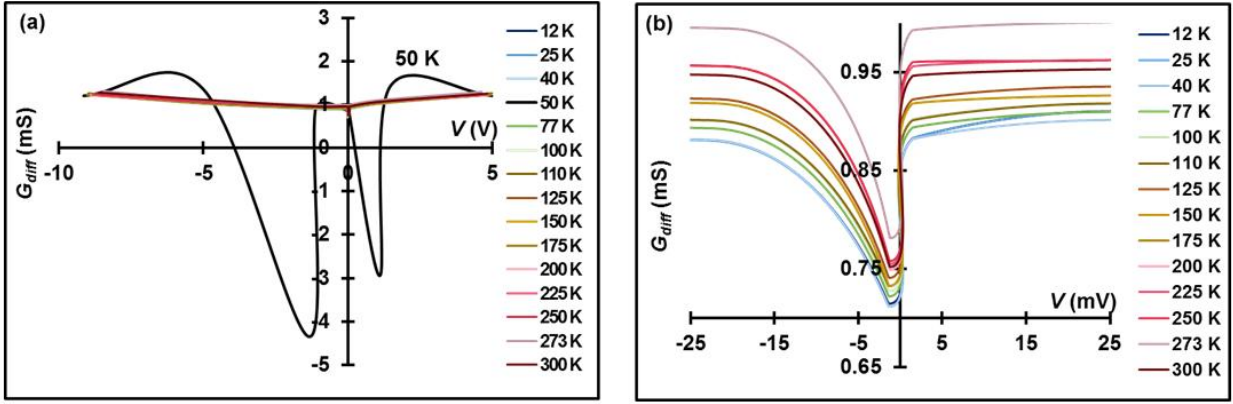

Fig. 1. Nonlocal $G_{diff}(V)$ data for an H-C-N fiber at high (a) and low voltages (b),

Analysis of Fig. 3 (Fig. 4b in [1]) suggests the existence of a chiral spin-triplet $p$-wave with two gap amplitudes: $\Delta_{V<0} \simeq 1.6$ eV located at $V \simeq -1.34$ eV and $\Delta_{V>0} \simeq 0.6$ eV located at $V \simeq 1.11$ eV, respectively. For voltages smaller than the SC gap at $T = 50$ K, the transport is dominated by Andreev reflections. The gap asymmetry is due to the charge imbalance that creates different rates at which the electron-like and hole-like quasiparticles are evacuated from the Andreev bound states and/or by the destruction of the chiral symmetry by the magnetic exchange field due to the itinerant FM introduced in the system by octane with its freely moving protons ($H^+$) on graphite's interfaces. We also notice the Fano-line shape of $G_{diff}$. The Fano line is a manifestation of coexisting polarons and Fermi particles in a superlattice of quantum wires. Note that in the stripe scenario for HTSC, a Fermi liquid coexists with an incommensurate 1D charge density wave forming a multi-gap SC near a Lifshitz transition where an amplification in the critical temperature $T_c$ is driven by Fano resonances involving different condensates. Thus, $G_{diff}$ data shows that SC correlations might be established in the H-C-N fibers below $T_c \sim 50$ K. This value is close to the mean-field $T_c$ for SC correlations in the metallic-H multilayer graphene or in highly oriented pyrolytic graphite, $T_c \sim 60$ K [12].

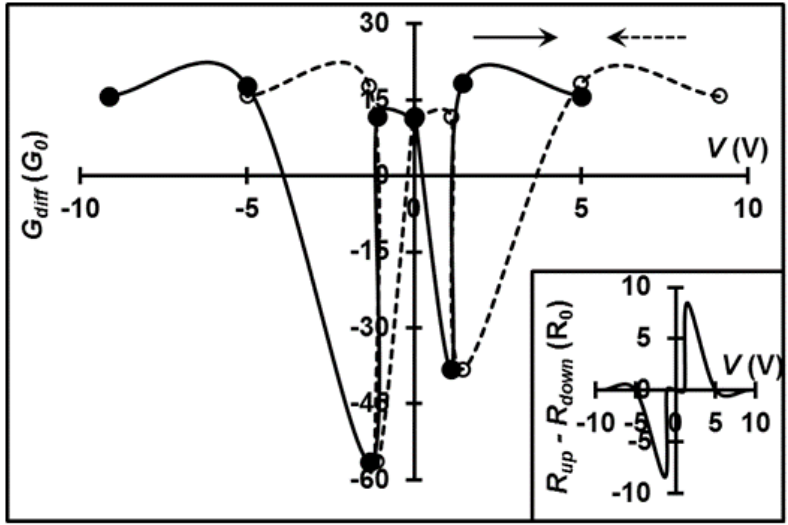

Fig. 2. Nonlocal $G_{diff}(V)$ experimental data for the H-C-N fiber at $T = 50$ K. The shape of the plot suggests interference of chiral asymmetric Andreev edge states and crossed Andreev conversion.

The $\Delta(T)$ data was further analyzed within the framework of the mean-field theory. One unusual SC behavior is manifested in the flat-band (FB) nature of the excitation spectrum, where the group velocity goes to zero, $d\omega(k)/dk \to 0$ [13]. As the FB states are highly localized around certain spots in the system, the SC order parameter becomes strongly inhomogeneous. The $\Delta(T)$ dependence is the solution to a transcendental equation: $\Delta(T) = \Delta_{FB}\tanh[\Delta(T)/(2k_B T)]$. This $\Delta(T)$ dependence comes from the density functional theory approach for SC, where the extended Kohn-Sham equation is written in the form of the Bogoliubov-deGennes equation used in the conventional theory for the description of inhomogeneous SC [14]. At the transition, the condition for the FB energy gap $\Delta_{FB}$ translates as $\Delta(T_c) \to 0$. Taylor series expansion to the first order gives $T_c \simeq \Delta_{FB}/2k_B$. Our results for the H-C-N fiber are shown in Fig. 3a (Fig. 6a in [1]). The $\Delta(T)$ gap data for the H-C-N fiber was replotted with the constant $\Delta_{FB}$ taken as the average of all $\Delta(T)$ values except for the data point at $T = 50$ K, $\Delta_{FB} = \Delta_{av} \simeq 12$ meV (Fig. 3a or Fig. 6a in [1]). While indeed constant ($T$-independent) below 50 K, the outstanding feature is the linear $T$-dependence of $\Delta_{FB}(T)$ for $T > 50$ K (Fig.3b or Fig. 6b in [1]). In addition, the lower inset in Fig. 3b shows that a linear $\Delta_{FB}$ ($T > 50$ K) implies the divergence of $\Delta(T = 50$ K).

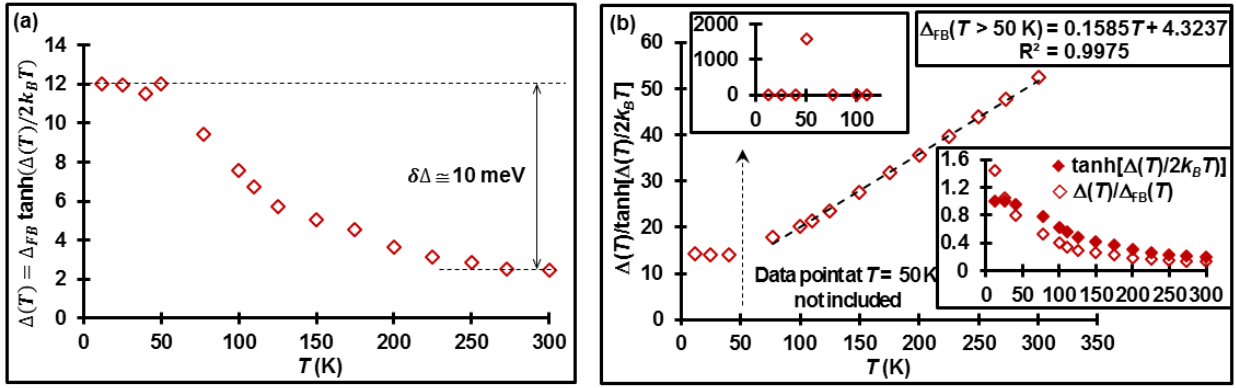

Fig. 3. a) The flat band $T$-dependence of the gap $\Delta(T)$ for a H-C-N fiber. The flat band parameter $\Delta_{FB}$ is the average of all $\Delta(T)$ values except for the data point at $T = 50$ K, $\Delta_{FB} = \Delta_{av} \simeq 12$ meV. b) The flat band parameter $\Delta_{FB}$ is constant below 50 K and linearly increases with $T$ above 50 K. The inset shows that a linear $\Delta_{FB}$ ($T > 50$ K) implies the divergence of $\Delta(T = 50$ K).

Notice that the conductance values are one order of magnitude larger than the quantum conductance $G_0 = ge^2/h \cong 77$ μS for a filling factor $g = 2$, where $e \cong -1.6 \times 10^{-19}$ C is the electron charge and $h \cong 6.62 \times 10^{-34}$ J·s is Planck's constant for quantum phenomena. Thus, at low $T$ there would be about 12 quantum conductance channels. Due to scattering of charges, in reality $G = G_0 \sum T_i \leq 12 G_0$, where $0 \leq T_i \leq 1$ are the transmission coefficients through channels. It might be interesting to compare our results with the ones known for conventional SC like Pb and Sn. The $T = 0$ K double gap for one layer of lead is $2\Delta_0(\text{Pb}) \cong 2.72$ meV, while for one layer of tin is $2\Delta_0(\text{Sn}) \cong 1.12$ meV. The $2\Delta/k_B T$ factor is 3.5 for Sn and 4.38 for Pb, respectively (see for instance [15]). The H-C-N fibers here have a gap ~12 larger, about the same ratio as $G/G_0$. Thus, $T_c$ would also need to be ~12 larger in the H-C-N fiber than in Sn ($T_c = 3.72$ K), i.e., $T_c \cong 12 \times 3.72/1.12 \cong 40$ K. Instead, we have found $T_c \cong 50$ K. This is because SC in Sn and Pb is electron-phonon mediated, while SC in the H-C-N fiber is a result of electron-electron (e-e) correlations, thus leading to higher $T_c$ and even HTSC.

Since we are dealing with quantum phenomena, we can therefore probe the quantum signature for the electron-hole exciton condensation in the H-C-N fiber by using the Heisenberg relationship: $\Delta x \cdot \Delta p_x \geq \hbar/2$. For graphene, it was found that the optimal distance between the H chains should be 2.5 Å [16]. Given the exciton energy $\Delta E \cong 1.6$ eV for the H-C-N fiber at $T_c \cong 50$ K, then $\Delta p_x \cong 2.1 \times 10^{-25}$ kg·m/s, thus by considering $\Delta x = 2.5$ Å, we find for the effective mass of the exciton $m^* \cong 0.1 m_e$, where $m_e \cong$ $9.1 \times 10^{-31}$ kg is the electronic mass. Interestingly, the effective mass $m^* \cong 0.1 m_e$ was also found to be the corresponding value in a magnetic field $B = 9$ T that was applied normal to the H-C-N fiber (see Table 2 in [1]). Recently, a photoinduced normal state in a HTSC showed a solid resemblance to the normal states under magnetic fields in equilibrium conditions [17]. Their approach is the reverse of ours [1] (see results depicted in Fig. 3 and discussion). There, we have found that a normal magnetic field of magnetic induction $B = 9$ T results in spin-triplet FM HTSC with a $T_c \sim 50$ K. Our result is close to the $T_c \sim 65.5$ K value found in [17] and the mean-field $T$ for SC correlations in the metallic-H multilayer graphene or in HOPG, $T_c \sim 60$ K [12]. At the same time, the exciton's energy ~1.6 eV (at $B = 0$ T) found in [1] is close the 1.55 eV energy of the photon used in [17]. It is possible that at the basis for these observations is a fundamental behavior of the electron-hole condensate that does not depend on the details of the HTSC system. Unlike the pairing of fermions in SC that is driven by antisymmetry, the pairing of particles and holes in the case of exciton condensation is driven by a more system-specific symmetry such as the geometric symmetry in the electronic double layers of GaAs and graphene [16]. We also notice that, whereas two-dimensional (2D) graphite is a zero-gap semiconductor, 3D graphite is semimetallic with a band overlap of 0.040 eV and a bandwidth along the Brillouin zone edge of 1.56 eV [18]. Within a many-body effects approach for graphene, electron-hole pair excitations amounting to a 1.6 eV energy would correspond to a dielectric constant $\varepsilon \sim 3 - 6$ [19].

For the unconventional HTSC found in H-C-N fibers [1], the role played by the H chains can be explained using Little's model for HTSC in organic materials [20]. The model considers a molecule consisting of two parts: a long chain called the "spine" in which electrons fill the various states and may or may not form a conducting system; and secondly, a series of arms or side chains attached to the spine. Under appropriate choice for the molecules which constitute the side chains, the virtual oscillation of charge in these side chains results in e-e correlations, some of which are SC correlations. Interestingly, even if the spine by itself is initially an insulator, the addition of side chains can increase the e-e interactions to the point where it becomes energetically favorable to enter the SC state by mixing in states of the conduction band. The spine thus transforms from the insulating or semiconducting state directly to the SC metallic state upon the addition of the side chains. In the case of the H-C-N fibers, the treatment with octane results in the free protonation of octane at graphite's interfaces [21]. The resulted superacidic H+ move freely without activation energy on the graphite surface giving rise to HTSC. Thus, when Little's model is applied to the H-C-N fiber, one can imagine that the spines are in the C planes with the H+ as the arms or side chains. The free H+ are shared by all C atoms in the plane to which the arms are connected to, thus the H+ mediate e-e correlations that need no input energy, i.e., they make the H-C-N fiber SC. A similar phenomenon was found in [22]. Thus, due to the off-diagonal long-range order that distinguished the SC from normal or insulating states, it is possible that Wigner's search for a quantum system to reproduce itself would lead to a SC as the only system where the probability for such an event would be nonzero. Also relevant is the observation that the Dirac-delta probability distribution of a particle on a benzene ring in one layer is paired with the probability distribution of a hole that is highly localized on the opposite benzene ring of the adjacent layer [16]. As pointed out there, molecular-scale excitonic condensates screened by electronic structure may be useful for the development of dissipationless molecular circuits and devices.

Moreover, the application of a high magnetic field on the H-C-N fiber leads to the formation of a special SC state known as the Fulde–Ferrell–Larkin–Ovchinnikov (FFLO) state that has been observed in other SC layered organic materials. The transition between the normal and the FFLO SC state is of second order. The inhomogeneous FFLO state results from the interaction of the SC condensate with a Zeeman field, the latter causing spin-splitting that in turn leads to the formation of Cooper pairs with a finite total momentum. For certain values of the Zeeman field, the ground state has an oscillating SC order parameter $\Delta(\vec{r}) \propto e^{i\vec{q}\cdot\vec{r}}$, where $\vec{q}$ is the total momentum of the Cooper pairs. There is an interplay between SC, Zeeman coupling, and Rashba spin-orbit effect in a 1D system and the stability of the FFLO state is protected by the spin-orbit interaction [23]. The existence of spin Seebeck effect due to the spin-orbit interaction was found from the linear temperature-dependence of the small SC gap $\Delta_S(T)$ for the H-C-N fiber, a quasi-1D system (Fig. 9 in [1]). In conjugated C-H structures with equal number of C atoms, the delocalized π electrons are like the conduction electrons in a metal. Moreover, the π electrons are like Cooper pairs in SC and their coherence leads to the Josephson tunnel effect [26]. Materials having as building blocks C and H atoms are practically small SCs, moreover ferromagnetic (FM), as we have previously found [1,27]. As a result of the competition between the magnetic and the SC orders, SC-FM bilayers show the formation of a quasi-1D FFLO-like that is due to Cooper pairs migrating from the SC into the FM. Adding to this a charge-density order, tricritical or tetracritical behavior is possible. While the tricritical point and its relation to the FFLO state can be revealed from the analysis of specific heat jumps in the system, recently, the FFLO state was found manifested as helical SC in noncentrosymmetric SCs [28]. In [1], we have concluded that that interpretation of our experimental data implies that the H-C-N fiber might contain chiral noncentrosymmetric tubular nanostructures. The noncentrosymmetric property results from mixing of the singlet *s*-wave and triplet *p*-wave states. We have evidence that in the H-C-N fiber the singlet and triplet states dominate over each other depending on the temperature range (results not yet published). There is competition and interplay between magnetic and SC order and electron viscosity can explain the observation of negative differential resistance and SC, as we have previously found [1,5,6]. Recently, a SC diode effect was found by reversing the direction of the magnetic field such that the SC acquires a finite resistance that is due to the free (unpaired) electrons flowing in the opposite direction to the SC current. The effect, which was attributed to the nonreciprocity of the Landau critical current for a metal-SC transition, can be used to probe the phase diagram for helical SC [28]. We have previously probed another manifestation of the SC diode effect in a bulk H-C-N material, where the behavior of the remanent magnetization was found to be consistent with the development of spin waves in a 2D Heisenberg model with a weak uniaxial anisotropy (see Fig. 6/left in [6], details to follow). At the same time, also being revealed is a unique manifestation of the interplay between SC and inversion breaking of the SC pairs (depairing). The observed phenomena are relevant to dissipationless electric circuits that again, are an attractive feature of molecular electronics. The negative current density also implies that the H-C-N fiber, as a granular SC vs. a homogeneously disordered SC, is more like a quantum *XY* spin glass than a disordered Bose liquid [24]. We have found spin glass behavior in a bulk graphitic sample [6], as well as hydrogenated graphitic samples (results not yet published).

In [1], we have also estimated the temperature for the Berezinskii–Kosterlitz–Thouless (BKT) transition to $T_{BKT} \sim 50$ K. Subsequently, we have better estimated the $T_{BKT}$ value. As known, the BKT transition is only relevant for exciton gaps $\Delta_{exciton} \sim 0.1 E_F$. In [1], the *T*-dependent conductivity was fit to Koike formula [25] that contains a term quantifying the Kondo effect owning to the presence of localized spins in the heat-treated, moreover hydrogenated C fibers. Using the value found for the Fermi temperature $T_F \cong 397$ K, then the Fermi energy is estimated to $E_F \cong 34$ meV from which we find $T_{BKT} \cong 43$ K. The Josephson coupling between two adjacent islands in the H-C-N fiber can be estimated as: $E_J \cong k_B T_{BKT} \cong 3.7$ meV, which is close to the zero-temperature small SC gap $\Delta_S \cong 4.6$ meV that we have found. Interestingly, $E_J/\Delta_S \approx T_{BKT}/T_c$, with the critical temperature for the excitonic condensate as found in [1], $T_c \cong 50$ K. That is, the H-C-N fiber is a granular SC material made out of large organic molecules in which the coherence of π electrons can lead the formation of Cooper SC pairs and to the Josephson tunnel effect. It is important to notice that a strong magnetic field does not destroy the pair correlation of π electrons [26]. Significantly, it was noticed that all the most important biologically active molecules contain conjugated systems. The pair correlation of the shared electrons, which leads to the gap in the spectrum, ensures stability and is analogous to the one observed in SC materials. Thus, the long-range order of correlated electrons plays an important role in the mechanism responsible for the coupling and transfer of excitations in biochemical materials.

Last but not least: The hydrogenation of graphitic layers after treatment with an alkane results in shape changing due to *sp²* to *sp³* bond conversion. This property of H-modified graphite, in particular graphene, to behave as a shape-changing membrane is essential for its relation to amino acids and to the SC transition at $T_c \sim 250 - 350$ K predicted by A. Salam that emphasizes the role of chirality in the origin of life and that we have previously revealed in the H-C-N system [27]. Thus, the several unexpected SC phenomena observed in the H-C-N fibers might be relevant to the important study of SC effects in biomatter to benefit humankind at its most crucial time.

**Acknowledgments**

The experimental part of this work was supported by The Air Force Office of Scientific Research (AFOSR) for the LRIR #14RQ08COR & LRIR #18RQCOR100 and the Air Force Research Laboratory within the Aerospace Systems Directorate (AFRL/RQ). N. Gheorghiu acknowledges G.Y. Panasyuk for his continuous support and inspiration that made possible this publication.